\documentclass[runningheads]{llncs}
\usepackage{cite}
\usepackage{amsmath,amssymb,amsfonts}
\usepackage{graphicx}
\usepackage{subfigure}
\usepackage{textcomp}
\usepackage{xcolor}
\usepackage{booktabs}
\usepackage{multirow}
\usepackage{amstext}
\usepackage{bm}
\usepackage{algorithm}
\usepackage{algpseudocode}
\usepackage[misc]{ifsym}
\begin{document}
\title{DINE: A Framework for Deep Incomplete Network Embedding}
\author{Ke Hou\inst{1} \and
Jiaying Liu\inst{1} \and
Yin Peng\inst{1}\and
Bo Xu\inst{1}\textsuperscript{(\Letter)}\and
Ivan Lee\inst{2}\and
Feng Xia\inst{3}}

\authorrunning{K. Hou et al.}
\institute{Key Laboratory for Ubiquitous Network and Service Software of Liaoning Province, School of Software, Dalian University of Technology, Dalian 116620, China\\
\email{boxu@dlut.edu.cn}\and
School of Information Technology and Mathematical Sciences, University of South Australia, Adelaide, SA 5095, Australia
\and School of Science, Engineering and Information Technology, Federation University, Australia}
\maketitle              % typeset the header of the contribution
\begin{abstract}
Network representation learning (NRL) plays a vital role in a variety of tasks such as node classification and link prediction. It aims to learn low-dimensional vector representations for nodes based on network structures or node attributes. While embedding techniques on complete networks have been intensively studied, in real-world applications, it is still a challenging task to collect complete networks. To bridge the gap, in this paper, we propose a Deep Incomplete Network Embedding method, namely DINE. Specifically, we first complete the missing part including both nodes and edges in a partially observable network by using the expectation-maximization framework. To improve the embedding performance, we consider both network structures and node attributes to learn node representations. Empirically, we evaluate DINE over three networks on multi-label classification and link prediction tasks. The results demonstrate the superiority of our proposed approach compared against state-of-the-art baselines.

\end{abstract}
\section{Introduction}
Information networks (e.g. citation networks, social networks, biological networks) contain different types of entities and intricate relations. Analyzing these networks plays an important role in many disciplines\cite{zhang2018network}. For example, in citation networks, we can find influential entities (i.e., scholars, papers) by calculating the importance of vertices\cite{cai2019scholarly,bai2017role}. In social networks, clustering users into communities is useful for recommendation\cite{wang2019sustainable,xia2014socially}. In biological networks, measuring the similarity between proteins helps us better understand protein interactions\cite{xu2018protein}. However, with the increase of entities and relations in real-world networks, it is challenging to explore the underlying network structures.

\begin{figure*}
	\centering
	\includegraphics[height=5.5cm,width=12cm]{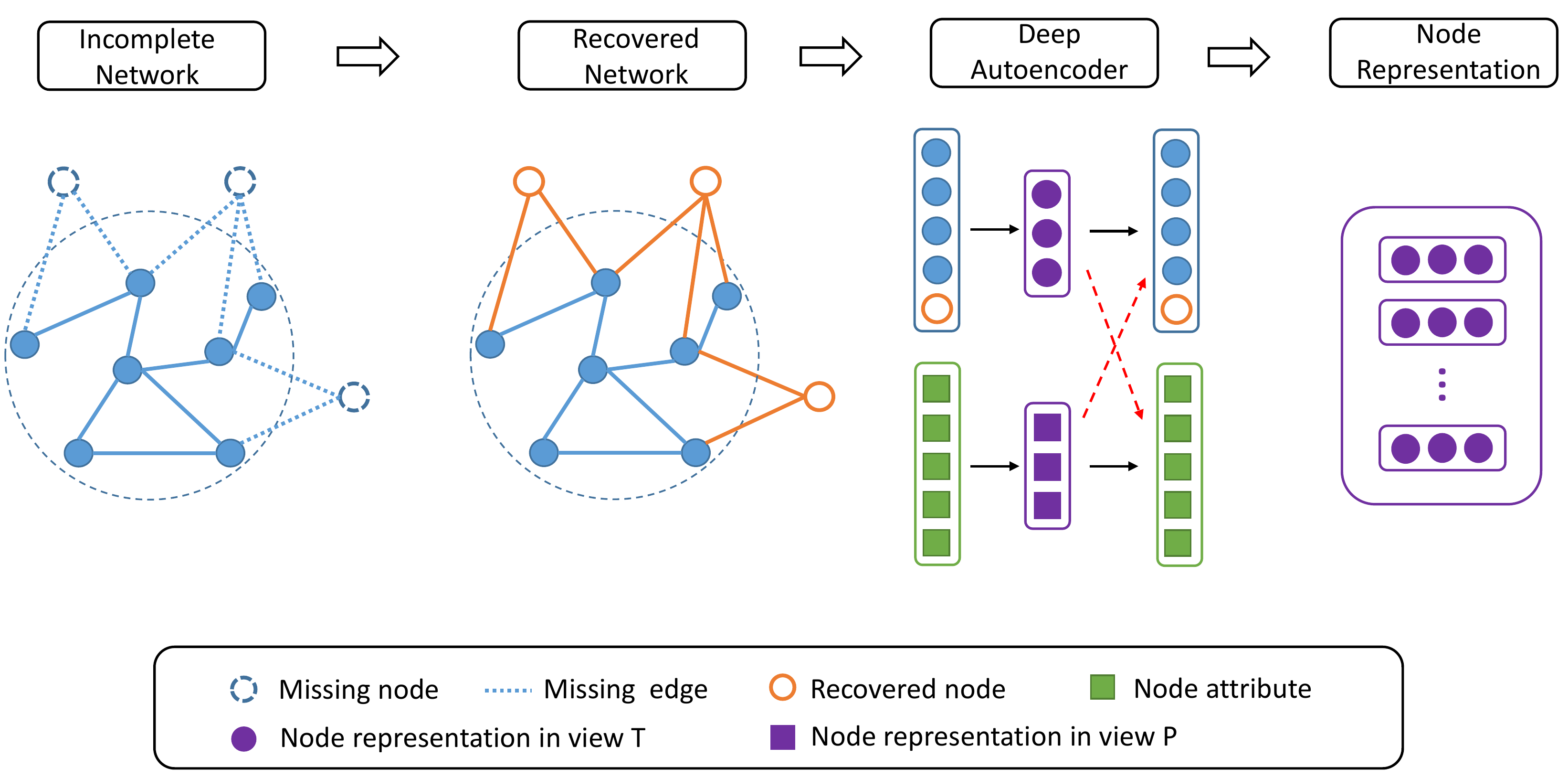}
	\caption{The overview of DINE framework.}
	\label{1}
\end{figure*}

To find an efficient way to model networks, researchers focus on network representation learning (NRL). NRL aims to learn latent, low-dimensional representations for nodes, with preserving not only network topologies but also node contents. Perozzi et al.~\cite{perozzi2014deepwalk} first combine NRL with skip-gram and propose Deepwalk, which lays a solid foundation for future development in this area. Recent advances in NRL have witnessed powerful representations abilities such as DeepGL\cite{rossi2018deep}, DANE\cite{gao2018deep}. Taking advantage of its powerful representation ability to model complex structures, NRL achieves significant performance in downstream tasks such as node classification\cite{zhu2007combining, bhagat2011node}, link prediction\cite{lu2011link,gao2011temporal}, and network visualization\cite{tang2016visualizing}.

In practice, many real-world networks are incomplete\cite{kossinets2006effects}, which further complicates the embedding process. For example, citation networks are usually incomplete because it is impossible for academic search engines to collect every paper. In biological networks, there exist a huge amount of undiscovered links because of the complexity of gene expression. Analyzing incomplete network makes a deviation because only a part of links are observed, which alters our estimates of network-level statistics. To fill this gap, researchers focus on network completion problem, which makes use of observed connectivity patterns to infer the missing part. However, existing studies only pay attention to missing links inference\cite{clauset2008hierarchical,guimera2009missing}, few of them focus on the incomplete networks with both missing nodes and edges\cite{kim2011network}.

To solve the problem, we present a new framework, named DINE for deep incomplete network embedding. DINE intelligently combines network completion and NRL into a unified framework. As shown in Fig.~\ref{1}, DINE contains two pivotal steps, including network recovery and network embedding. Specially, we first capture the connectivity patterns from the partially observable network and fit the generative graphs model to estimate missing components. To model the network more accurately, we consider both network structures and node attributes to learn the representations of the recovered network by using a deep autoencoder. Finally, we empirically verify the performance of the proposed framework on three real-world networks. Experimental results illustrate the significant representation ability of DINE in partially observable networks. Our main contributions can be concluded as follows:
\begin{itemize}
\item[(1)] We present a new framework, namely DINE, for deep incomplete network embedding. DINE intelligently combines network completion and NRL into a unified framework, which provides an effective solution for data missing.
\item[(2)] DINE considers not only topology structure but also node attributes for embedding. It can accurately and effectively model node proximity and underlying structure in the joint space.
\item[(3)] We extensively validate the framework on three real-world networks through multi-label classification and link prediction tasks. The results demonstrate the superiority of our proposed approach compared with state-of-the-art baselines.
\end{itemize}

The remainder of this paper is organized as follows. Section 2 summarizes related work. In section 3, we focus on problem definition. Section 4 introduces the implementation details of the proposed framework. Experimental results are provided in section 5. Finally, we conclude this work in section 6.
\section{Related Work}
The framework we proposed in this paper is related to two areas of research, including network completion and NRL techniques. 

\subsection{Network Completion}
Network completion deals with the problem of inferring missing nodes and edges in networks. Network completion is similar to matrix completion \cite{keshavan2010matrix}, which aims to complete the matrix with elements missing. However, network completion is more arduous because of network diversity. For missing edges, it is an attractive way to recover the original network by calculating node similarity. Another way to complete missing edges is considering shared node neighbors\cite{buccafurri2015discovering}. In cases where both nodes and edges are missing, we can utilize a generative graphs model named KronFit\cite{leskovec2010kronecker} to generate complete networks whose structures are similar to real-world networks. Kim et al.\cite{kim2011network} combine expectation-maximization into KronFit, and propose a powerful algorithm KronEM, which is more effective for recovering the network.

\begin{table}[h]\renewcommand{\arraystretch}{1}
	\centering
	\caption{The description of notations }
	\label{Description of notations}
	\begin{tabular}{cl}
		\toprule
		Notation & \multicolumn{1}{c}{Description} \\
		\midrule
		$N_{G}$ & number of nodes in the complete network\\
		$\mathit{{N^{\prime}}}$ & number of nodes in the incomplete network \\
		$\mathit{{N_{M}}}$ & number of missing nodes \\
		$\mathit{{N_{R}}}$ & number of recovered nodes \\
		$xt$ & input of network structure view $T$ \\
		$xp$ & input of node attribute view $P$ \\
		$x\hat t^t$& reconstruction output of $xt$ by self-view \\
		$x\hat t^p$& reconstruction output of $xt$ by cross-view \\
		$x\hat p^t$& reconstruction output of $xp$ by self-view \\
		$x\hat p^p$& reconstruction output of $xp$ by cross-view \\
		$K$& number of encoding layers\\
		$yt^{(K)}$ & representation in network structure view $T$\\
		$yp^{(K)}$ & representation in node attribute view $P$\\
		$\alpha$ & balance reconstruction errors of self-view and cross-view\\
		$\beta$ & balance reconstruction errors of $G^{\prime}$ and $G_{M}$\\
		\bottomrule
	\end{tabular}
\end{table}

\subsection{Network Representation Learning}
NRL aims to embed each node in the network into a low-dimensional representation. Existing NRL algorithms can be divided into four categories. The first category is matrix factorization based methods. They first represent the connections between network vertices and use matrix factorization to obtain representations. IsoMAP\cite{tenenbaum2000global} constructs an affinity network by feature vectors. It represents nodes with the solved leading eigenvectors. The second category is random walk based methods. DeepWalk\cite{perozzi2014deepwalk} utilizes random walk to learn structural information and uses skip-gram to obtain the representations. Node2vec\cite{grover2016node2vec} changes the strategy of random walk to capture a more global structure. The third category is edge modeling based methods. They utilize node-node connections to learn node representations directly. LINE\cite{tang2015line} uses first-order proximity and second-order proximity to obtain local and global structure information. The fourth category is deep learning based methods. They could extract highly non-linear structure automatically by using deep learning techniques. SDNE\cite{wang2016structural} preserves first and second order proximities for highly non-linear structures via a deep autoencoder.

\section{Preliminary}
In this section, we first describe the notations used in this paper. We then formalize the problem of network embedding in an incomplete network.
\subsection{Notations}
We denote the complete network as $\mathit{G}=\left(\mathit{V},\mathit{A},\mathit{P}\right)$, where $\mathit{V}=\left\lbrace\mathit{v}_{1},\mathit{v}_{2},...,\mathit{v}_{\left| V\right| }\right\rbrace$ indicates the nodes in the network.
$\mathit{A}\in\mathbb{R}^{\left|V\right|\times\left|V\right|}$ represents the adjacency matrix and $\mathit{P}\in\mathbb{R}^{\left|V\right|\times\left|P\right|}$ denotes the node attribute matrix, where $\left|V\right|$ and $\left|P\right|$ represent the dimension of adjacency matrix and node attributes, respectively. Similarly, we define the incomplete network, the missing network, and recovered network as $\mathit{G^{\prime}}=\left(\mathit{V}^{\prime},\mathit{A}^{\prime},\mathit{P}^{\prime}\right)$, $\mathit{G_{M}}=\left(\mathit{V}_{M},\mathit{A}_{M},\mathit{P}_{M}\right)$,  $\mathit{G_{R}}=\left(\mathit{V}_{R},\mathit{A}_{R},\mathit{P}_{R}\right)$, respectively. Table~\ref{Description of notations} lists the meaning of the notations mainly used in this paper.

\subsection{Problem Formulation}
The purpose of network completion is to infer the missing part of the incomplete network, how to infer the missing network $\mathit{G_{M}}$ from the observable network $\mathit{G^{\prime}}$ is crucial to the problem. If we use adjacency matrices to represent the network, then the network completion problem can be transformed into matrix completion problem. In general, classical matrix completion problem is to determine the value (0 or 1) of elements in the missing part in a binary matrix. In this paper, we assume that the number of missing nodes is known. If not, the standard methods for estimating the size of hidden (missing) populations can solve this problem\cite{mccormick2010many}.

Although network recovering helps in representing the incomplete network, there are some problems in the representation learning process. On the one hand, many network representation methods are shallow models. Network completeness is essential for extracting local or global topology information. On the other hand, most methods can't capture non-linear relations between nodes\cite{wang2016structural}. Thus, we need to consider not only topology information for non-linear relations but also node contents such as node attributes. Besides, $\mathit{A}^{\prime}$ and $\mathit{P}^{\prime}$ preserve the information of a network, which is used to represent the network in the joint space. Thus, nodes with similar topology structures or attributes will be closer in the representation dimension.

\section{Design of DINE}
In this section, we present a novel framework, namely DINE, to solve the problem of network embedding in incomplete networks. Our framework contains two crucial components, network recovery and network embedding. Firstly, we discuss how to recover the incomplete network. Then, we introduce the process of network representation learning,  which considers both topology information and node attributes.

\subsection{Recovery of Incomplete Network}
To recover the network with nodes and edges missing, we model the incomplete network with the Kronecker graphs model\cite{leskovec2010kronecker}. In detail, we use the incomplete network to fit the Kronecker graphs model in network structure and estimate the missing part, and then re-estimate model parameters. These two steps are iterated until the model parameters converge. Finally, we obtain the missing part of the network.

The purpose of the network completion is to find the most likely structure of the missing part $\mathit{G_{M}}$. We connect the incomplete network and the missing network by network generation parameters $\Theta$. Let $\sigma $ denote the mapping among nodes in the recovered network, incomplete network, and missing network. The mapping $\sigma $ indicates a permutation of set $\left\lbrace 1,... ,N_{G}\right\rbrace$. The first $\mathit{N^{\prime}}$ elements of $\sigma $ map the nodes of $ G_{R} $ to the incomplete network and the remaining $\mathit{N_{M}}$ elements of $\sigma $ map the nodes of missing part $\mathit{G_{M}}$. The likelihood $P\left( G^{\prime},G_{M},\sigma| \Theta\right) $ can be represented as:
\begin{small}
	\begin{equation}
	P\left( G^{\prime},G_{M},\sigma| \Theta\right) = \prod_{a_{uv}=1}\left[\Theta^{k}\right]_{\sigma\left( u\right) \sigma\left( v\right) }
	\prod_{a_{uv}=0}\left( 1-\left[\Theta^{k}\right]_{\sigma\left( u\right) \sigma\left( v\right) }\right)
	\end{equation}
\end{small}
where $\Theta^{k} $ is the adjacency matrix generated by model parameters $\Theta$. $ \left[\Theta^{k}\right]_{\sigma\left( u\right) \sigma\left( v\right)}$ denotes the $ \left(\sigma\left( u\right),\sigma\left( v\right)\right)$-th element of matrix $\Theta^{k}$. $ a_{uv} $ is the $ \left( u,v\right)$-th element of $A_{R}$, which is the the adjacency matrix of the recovered network.

Next, we consider the edges in the missing part and $\sigma$ as the latent variables. E-step is to sample the missing part and permutation. M-step aims to optimize the parameters $\Theta$ by stochastic gradient descent process. Then we iterate E-step and M-step until parameters $\Theta$ converge. The steps could be described as:

$ E\verb|-|step: $
\begin{equation}
( G_{M}^{( t) },\sigma ^{( t) })  \sim P( G_{M},\sigma|G^{\prime},\Theta^{( t) })
\end{equation}

$ M\verb|-|step: $
\begin{equation}
\Theta ^{( t+1) }=\mathop{\arg\max}_{\Theta\in( 0,1) ^{2}} \mathbb{E} [P( G_{M}^{( t) },\sigma ^{( t) },G^{\prime}|\Theta )].
\end{equation}

In detail, we first initialize model parameters $\Theta$ and generate a stochastic network. Then we sample the missing part $G_{M}$ and node mapping $\sigma$ by Gibbs sampling, which can be considered to recover the missing part of the network. Besides, we optimize the model parameters $\Theta$ and iterate the above steps until the parameters converge. Finally, we obtain the most likely instances of the missing part and node mapping.

\subsection{Recovered Network Embedding MVC-DNER}
\begin{figure}
	\centering
	\includegraphics[height=6cm,width=9cm]{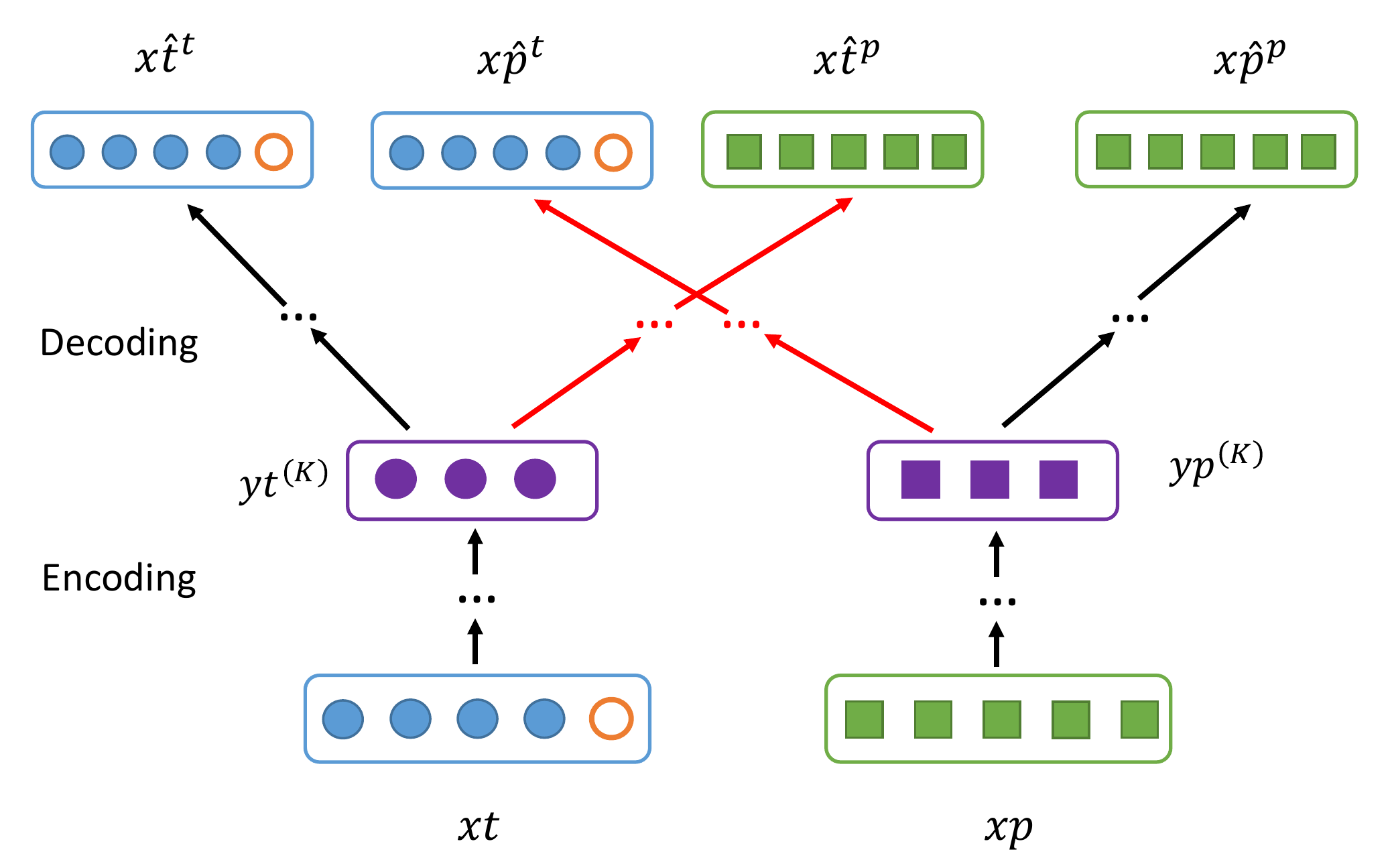}
	\caption{Deep autoencoder MVC-DNER.}
	\label{2}
\end{figure}

In terms of network representation, we consider not only network topology structure but also node attributes. Furthermore, inspired by MVC-DNE\cite{yang2017properties} which utilizes a deep autoencoder, we propose MVC-DNER to capture non-linear structures and node attributes in the recovered network.
Fig.~\ref{2} shows that the embedding part has network structures view $T$ and node attributes view $P$, which uses deep autoencoder to learn latent information in each view. 
\begin{table}[htbp]	\renewcommand{\arraystretch}{1.5}
	\renewcommand{\tabcolsep}{10.0pt}
	\centering
	\caption{Layer structures of MVC-DNER on three datasets }	
	\label{tab:1}       	
	\begin{tabular}{ccc}
		
		\toprule
		
		Dataset & layers in view V & layers in view P  \\
		
		\midrule
		
		Citeseer & $N_{R}$-500-128 & 3,703-600-128 \\
		
		DBLP & $N_{R}$-800-128 & 9,662-900-128 \\
		
		BlogCatalog &$N_{R}$-500-128  & 39-4 \\
		
		\bottomrule
		
	\end{tabular}	
\end{table}
We take the adjacency matrix $xt$ and the attribute matrix $xp$ of the recovered network as input. In the encoding process, input features of one view could encode some shared latent information reflecting the input of the other view. In the decoding process, latent representations in one view could reconstruct the input of another view. The loss function is defined as:
\begin{equation}
L(xt,xp;\theta)=L_{t}(xt,xp;\theta)+L_{p}(xt,xp;\theta)
\end{equation}
\begin{equation}
\begin{split}
L_{t}(xt,xp;\theta)=\beta\sum_{i=1}^{|V'|}((1-\alpha)\left\|xt_{i}-x\hat t_{i}^t\right\|_{2}^{2}+\alpha\left\|xt_{i}-x\hat t_{i}^p\right\|_{2}^{2})\\
+(1-\beta)\sum_{i=1}^{|V_{M}|}((1-\alpha)\left\|xt_{i}-x\hat t_{i}^t\right\|_{2}^{2}+\alpha\left\|xt_{i}-x\hat t_{i}^p\right\|_{2}^{2})
\end{split}
\end{equation}
\begin{equation}
\begin{split}
L_{p}(xt,xp;\theta)=\beta\sum_{i=1}^{|V'|}((1-\alpha)\left\|xp_{i}-x\hat p_{i}^t\right\|_{2}^{2}+\alpha\left\|xp_{i}-x\hat p_{i}^p\right\|_{2}^{2})\\
+(1-\beta)\sum_{i=1}^{|V_{M}|}((1-\alpha)\left\|xp_{i}-x\hat p_{i}^t\right\|_{2}^{2}+\alpha\left\|xp_{i}-x\hat p_{i}^p\right\|_{2}^{2})
\end{split}
\end{equation}
where $ x\hat t_{i}^t $ and $ x\hat t_{i}^p $ are the reconstruction outputs of $xt_{i}$. $ x\hat p_{i}^t $ and $ x\hat p_{i}^p $ are the reconstruction vectors of $xp_{i}$. $\alpha$ and $\beta$ are parameters to adjust the proportion of self-view and cross-view reconstruction errors, recovered nodes and observed nodes reconstruction errors, respectively. $\theta= \{W^{(l)},b^{(l)},\hat W^{(l)},\hat b^{(l)}\}_{l=1}^{K}$ denotes parameters including the weights $W$ and bias $b$ in the deep autoencoder.

The loss function is minimized by stochastic gradient descent. Thus, the learning representations preserve not only network structures information but also node attributes information.

\section{Experiments}
In this section, we evaluate our framework on three datasets through multi-label classification and link prediction tasks. We first introduce three datasets and baseline methods. Then we describe evaluation metrics and parameter settings of the methods.  Finally, we present the performance of DINE and compare it against state-of-the-art baselines.
\subsection{Datasets}
We use three datasets including two academic datasets (Citeseer\footnote{https://linqs.soe.ucsc.edu/data} and DBLP\footnote{https://www.aminer.cn/billboard/citation}) and a social dataset BlogCatalog\footnote{http://socialcomputing.asu.edu/datasets/BlogCatalog}.

\begin{itemize}
	\item[(1)] {\bfseries Citeseer} contains citation information of papers. In the citation network, each node represents a paper and edges reflect citation relationship. The citation network constructed by Citeseer contains 3,312 papers divided into six classes including Agents, AI, DB, IR, ML, and HCI. Besides, the attribute feature of each paper is a 3,703 dimensional binary vector based on the topics.
	\item[(2)] {\bfseries DBLP} is also a citation dataset which covers useful information on papers such as authors, year, publisher, and title. It provides open bibliographic information of major computer science journals and proceedings. We choose 8,192 papers from 10 research domains. We choose the title of papers as the attribute and use a 9,662 dimensional binary vector to represent the attribute feature.
	\item[(3)] {\bfseries BlogCatalog} is a social blog directory that manages bloggers and their blogs. We choose some data which contains 4,096 nodes, 38,983 edges, and 39 groups. Nodes represent bloggers and edges represent the friendship between bloggers. Besides, each blogger belongs to one or several groups based on interesting.
\end{itemize}

\subsection{Baseline Methods}
We use the following methods as our baseline methods. We choose four network representation learning methods based on matrix factorization, random walk, and deep learning, respectively.
\begin{itemize}
	\item[(1)] {\bfseries GF} \cite{ahmed2013distributed} is a matrix factorization based method. It relies on partitioning a graph to minimize the number of neighboring vertices. In addition, it preserves first order proximity and allows for linear scalability.
	\item[(2)] {\bfseries HOPE}\cite{ou2016asymmetric} is also based on matrix factorization. It provides an efficient way to preserve high-order proximities of large-scale graphs. It is also capable of capturing the asymmetric transitivity.
	\item[(3)] {\bfseries Node2vec}\cite{grover2016node2vec} is a shallow model. It designs a flexible neighbor sampling strategy based on Deepwalk. It can preserve both local structure and global structure to learn network representations.
	\item[(4)] {\bfseries SDNE}\cite{wang2016structural} is the first network representation learning method based on deep learning. The deep autoencoder captures the non-linear network structure. It also can preserve the local and global network structure.
\end{itemize}

\subsection{Parameter Settings}
Our framework consists of network completion and recovered network representation learning. In the network completion, the Kronecker parameter $\Theta$ is random initialization. The neural network structure of  MVC-DNER is listed in Table~\ref{tab:1}. We set the learning rate as 0.001. The mini-batch size of optimization is 50. The parameters for balancing the importance of self-view and cross-view $\alpha$ , recovered nodes and observed nodes $\beta$ are set to 0.5 and 0.8, respectively.

The parameter settings of these baseline methods including GF, HOPE, Node2vec, and SDNE follow a NRL survey\cite{goyal2018graph}. The learning rate of SGD is 0.0001, and max iterations are 5,000 in GF. The higher-order coefficient of HOPE is 0.01. In Node2vec, we set the window size as 10, the walk length as 40, walks per node as 40. The dimension of network learning representation is 128 for all methods.
\begin{table}[ht]\renewcommand{\arraystretch}{1} %控制行高
	\renewcommand{\tabcolsep}{10.0pt}
	\centering
	\caption{Multi-label classification results(macro-F1) on two datasets with the portion of missing nodes}
	\label{3}
	\begin{tabular}{ccccccc}
		\toprule
		\toprule
		\multirow{1}{1.5cm}{Datasets}&$M_{r}$&GF&HOPE&Node2vec&SDNE&DINE\cr
		\midrule
		\multirow{6}{1.5cm}{Citeseer}
		~& 0.05 & 0.253&0.265&0.431&0.366&$\bm{0.642}$\cr
		~ & 0.10  &0.231&0.253&0.423&0.365&$\bm{0.636}$\cr
		~ & 0.15 & 0.202&0.257&0.412&0.358&$\bm{0.627}$\cr
		~ & 0.20 &0.203&0.244&0.419&0.352&$\bm{0.629}$\cr
		~ & 0.25 & 0.235&0.249&0.403&0.346&$\bm{0.617}$\cr
		~ & 0.30 &0.225&0.243&0.388&0.335&$\bm{0.614}$\cr
		\hline
		\multirow{6}{1.5cm}{DBLP}
		~& 0.05 &0.579&0.575&0.582&0.585&$\bm{0.595}$\cr
		~ & 0.10  &0.575&0.574&0.594&0.558&$\bm{0.601}$\cr
		~ & 0.15 &0.541&0.534&0.584&0.571&$\bm{0.594}$\cr
		~ & 0.20 &0.576&0.587&0.564&0.566&$\bm{0.591}$\cr
		~ & 0.25 &0.574&0.580&0.561&0.577&$\bm{0.593}$\cr
		~ & 0.30 & 0.572&0.571&0.558&0.573&$\bm{0.590}$\cr
		\bottomrule
		\bottomrule
	\end{tabular}
\end{table}
\subsection{Experimental Results}

{\bfseries Multi-label classification}. We aim to learn representations in an incomplete network. To achieve this goal, we need to remove 5\%-30\% nodes and the corresponding edges.  We first learn the representations based on remaining nodes and take the representations as the input of the classification model. Then we divide the labeled nodes into training set and testing set. The portion ratio of training nodes varies from 10\% to 90\%. We use macro-F1 to evaluate the performance of the classification model. Besides, the experiment runs 10 times, and we take the average of results as the final results. Table~\ref{3} lists the classification results for each method, where $M_{r}$ is the portion ratio of missing nodes.

From the table, we can observe that the performance of DINE is better than any other baseline methods, especially in the Citeseer dataset. The performances of these methods gradually become worse as the portion of missing nodes increasing. Besides, two matrix factorization methods have terrible performance in Citeseer. Most methods have a relatively better performance in the DBLP dataset.

\begin{figure*}[htbp]
	\centering
	\subfigure[Citeseer]{
		\begin{minipage}[t]{0.33\linewidth}
			\centering
			\includegraphics[width=4cm]{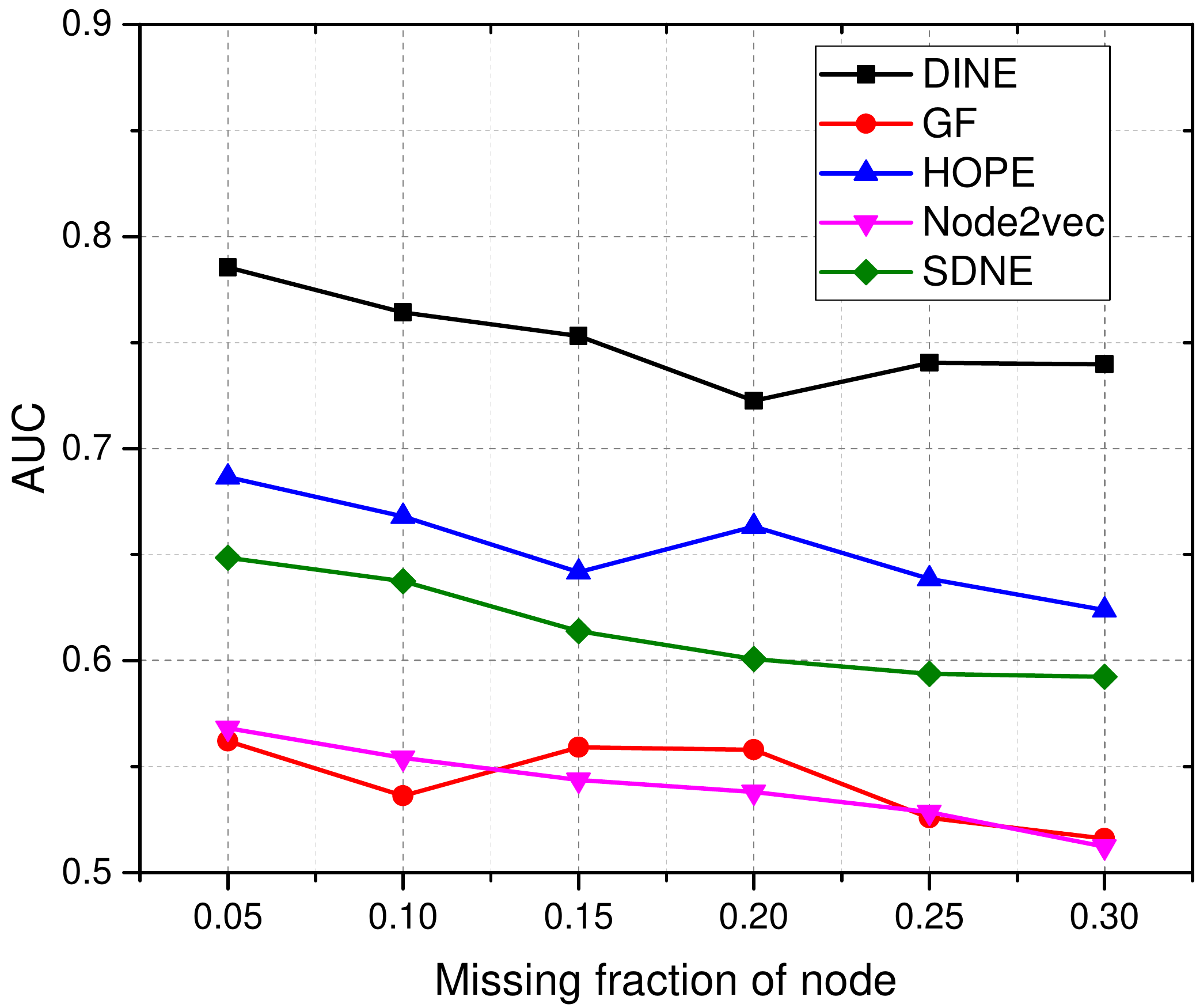}
			%\caption{Citeseer}
		\end{minipage}%
	}%
	\subfigure[DBLP]{
		\begin{minipage}[t]{0.33\linewidth}
			\centering
			\includegraphics[width=4cm]{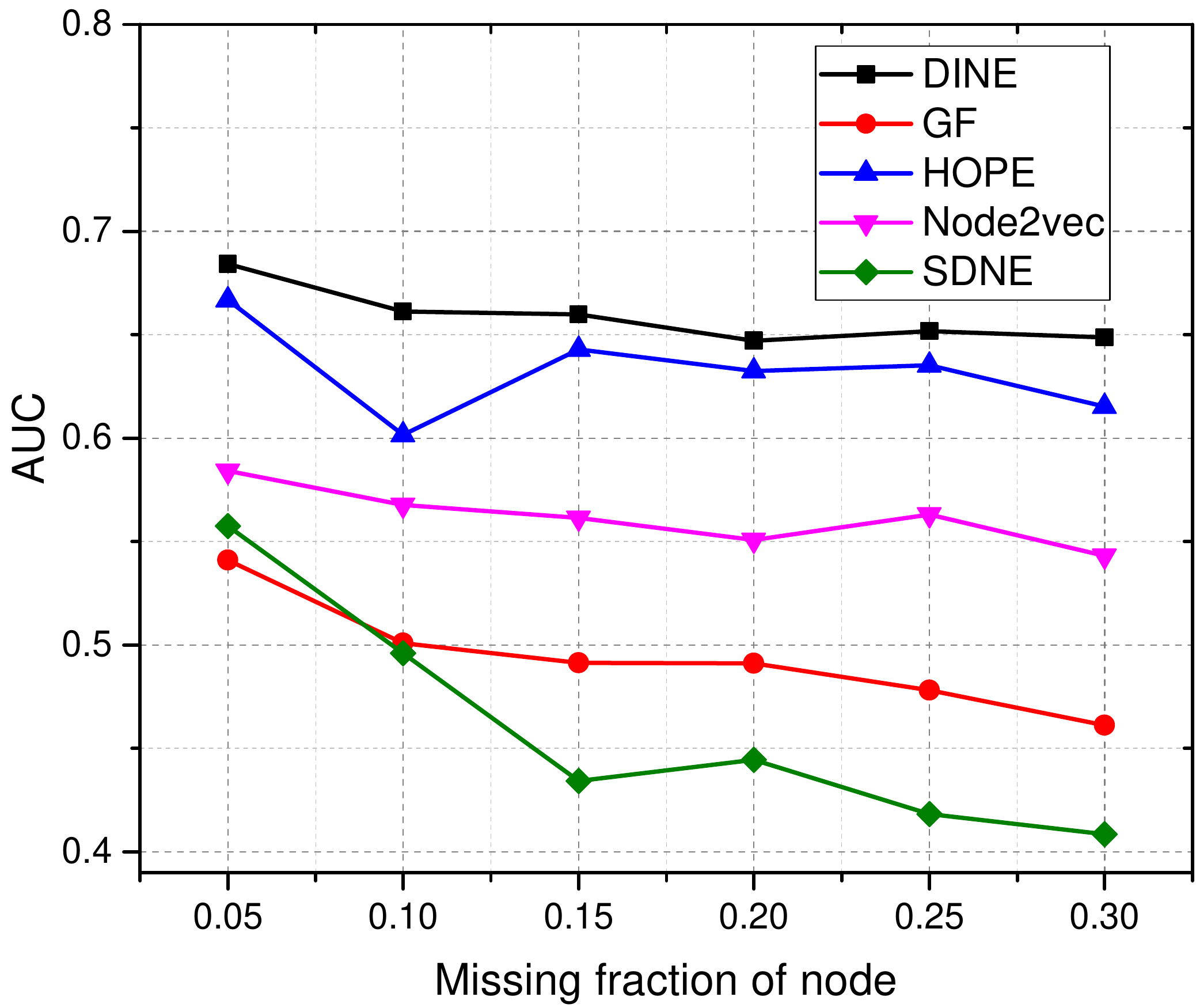}
			%\caption{DBLP}
		\end{minipage}%
	}%
	\subfigure[BlogCatalog]{
		\begin{minipage}[t]{0.33\linewidth}
			\centering
			\includegraphics[width=4cm]{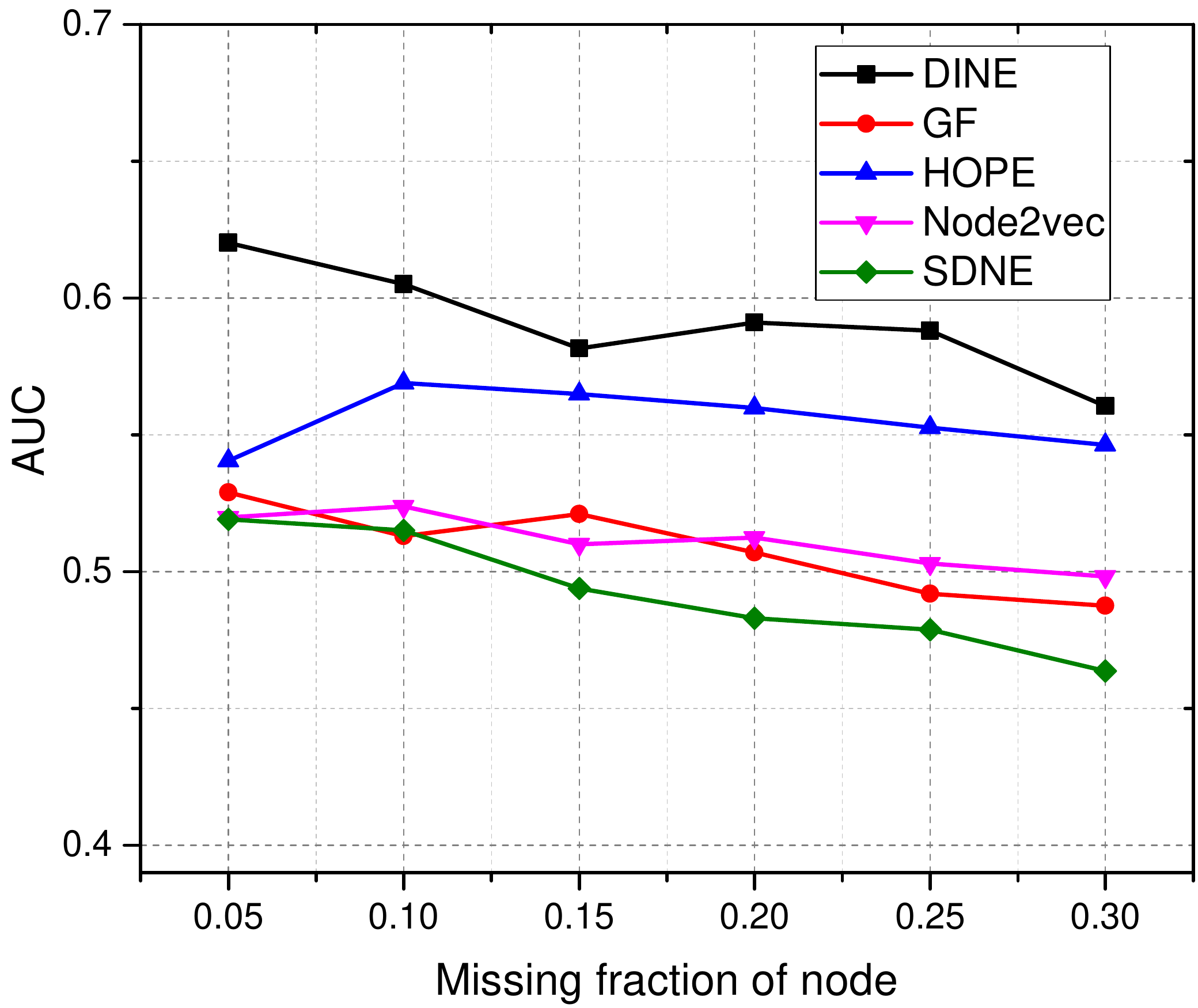}
			%\caption{BlogCatalog}
		\end{minipage}
	}%
	\centering
	
	\caption{Link prediction results (AUC) on three datasets with the portion of missing nodes.}
	\label{fig3}
\end{figure*}

{\bfseries Link prediction}. Similar to the task of multi-label classification, we also remove partial nodes and the corresponding edges. Then we remove 20\% edges of the remainder network as links for prediction and consider them as positive samples. Besides, we randomly select unconnected node pairs as negative samples. The number of negative samples is the same as positive samples. From the results of link prediction presented in Fig.~\ref{fig3}, we can see that DINE achieves significant improvements in AUC over the baselines in all datasets. As the portion of missing nodes increasing, the performances of these methods have a downward trend.

\section{Conclusion}
In this paper, we have presented a framework named DINE, which aims to learn node representations in incomplete networks. The framework is divided into two parts: network completion and recovered network representation learning. Specifically, we recover the missing part of the incomplete network based on the combination of EM approach and Kronecker graphs model. After recovering the incomplete network, we propose an algorithm named MVC-DNER to learn node representations for the recovered network. MVC-DNER uses the deep autoencoder to learn representations, which preserves both network structures and node attributes. Experimental results on three real-world network datasets show the significant performance of our proposed method. The future work is primarily on extending DINE to heterogeneous networks containing different types of nodes and edges.

\bibliographystyle{splncs04}
\bibliography{ref}
%Author, F.: Article title. Journal \textbf{2}(5), 99--110 (2016)

%\end{thebibliography}
\end{document}